\documentclass[11pt,preprint]{aastex}
\usepackage{epsfig}
 
\newcommand       \AU           {\,{\rm AU}}          
\newcommand       \cm           {\,{\rm cm}}

\newcommand	  \g		{\,{\rm g}}
\newcommand       \K            {\,{\rm K}}
\newcommand	  \pc		{\,{\rm pc}}

\newcommand	  \yr		{\,{\rm yr}}

\newcommand	  \Myr		{\,{\rm Myr}}

\newcommand       \simlt        {\lesssim}
\newcommand       \simgt        {\gtrsim}
\newcommand       \gtsim        {\gtrsim}

\newcommand       \mum          {\,{\rm \mu m}}

\newcommand	  \Teff	        {T_{\rm eff}}

\newcommand	  \hra	        {{\rm HR\,4796A}}

\newcommand	  \hda	        {{\rm HD\,141569A}}

\newcommand	  \amin	        {a_{\rm min}}
\newcommand	  \amax	        {a_{\rm max}}
\newcommand	  \rin	        {r_{\rm in}}
\newcommand	  \rout	        {r_{\rm out}}
\newcommand	  \msil         {m_{\rm sil}} 
\newcommand	  \mcarb        {m_{\rm carb}} 
\newcommand	  \mice         {m_{\rm ice}} 
 
\newcommand	  \md           {m_{\rm d}}
\newcommand	  \rp           {r_{\rm p}}

\newcommand	  \Pice         {P^{\prime}}
\newcommand	  \sigmap       {\sigma_{\rm p}}
\newcommand	  \sigmar       {\sigma(r)}

\newcommand       \msun         {m_\odot}

\newcommand       \mstar        {m_\star}

\newcommand       \mearth       {\,{m_\oplus}}

\newcommand       \eri          {\epsilon\ {\rm Eri}}

\newcommand       \chisq        {\chi^2/N_{\rm obs}}

\newcommand       \simali       {\sim\,}
\newcommand{\figwidth}{6.0in}

\countdef\decade=200
\advance\decade by \year
\countdef\hours=201
\advance\hours by \time
\divide\hours by 60
\countdef\mins=202
\advance\mins by \hours
\multiply\mins by 60
\multiply\hours by 100
\countdef\miltime=203
\advance\miltime by \hours
\advance\miltime by \time
\advance\miltime by -\mins

\shorttitle{The $\eri$ Dust Disk}
\shortauthors{Li, Lunine, \& Bendo}

\begin{document}

\title{
 \vspace*{-2.0em}
 {\normalsize\rm To appear in {\it The Astrophysical Journal Letters}, 
 the 2003-November-20th issue}\\
 \vspace*{1.0em}
Modeling the Infrared Emission from the $\epsilon$ Eridani Disk
	 }
\author{Aigen Li, J.I. Lunine, and G.J. Bendo}
\affil{Theoretical Astrophysics Program,
        Steward Observatory and 
        Lunar and Planetary Laboratory,  
        University of Arizona, Tucson, AZ 85721;\\
        {\sf agli@lpl.arizona.edu, jlunine@lpl.arizona.edu,
             gbendo@as.arizona.edu}}

\begin{abstract}
We model the infrared (IR) emission from the ring-like dust 
disk around the main-sequence (MS) star $\epsilon$ Eridani, 
a young analog to our solar system, in terms of a porous dust 
model previously developed for the extended wedge-shaped disk 
around the MS star $\beta$ Pictoris and the sharply truncated 
ring-like disks around the Herbig Ae/Be stars $\hra$ and $\hda$.
It is shown that the porous dust model with a porosity of $\simali$90\%
is also successful in reproducing the IR to submillimeter dust emission 
spectral energy distribution as well as the 850$\mum$ flux radial 
profile of the dust ring around the more evolved MS star $\eri$.
Predictions are made for future {\it SIRTF} observations which 
may allow a direct test of the porous dust model.
\end{abstract}

\section{Introduction\label{sec:intro}}
Together with Vega, Fomalhaut ($\alpha$ PsA), and $\beta$ Pictoris, 
the nearby (distance to the Earth $d \approx 3.2\pc$) 
low-mass ($\mstar \approx 0.8\msun$) K2V main-sequence (MS) star
$\epsilon$ Eridani constitutes the ``Big-Four'' family of
prototypical ``Vega-excess'' stars, which are MS stars with
infrared (IR) radiation in excess of what is expected 
from their stellar photospheres (Gillett 1986).
The IR excess associated with ``Vega-excess'' stars is generally
attributed to the thermal emission of dust grains orbiting 
the central star in the form of a disk or ring
and heated by its stellar radiation.
The dust is thought to be ``second generational'' in nature
rather than ``primordial''; i.e., the dust is considered as 
{\it debris} generated by collisions of larger objects such as 
cometary and/or asteroidal bodies rather than {\it remnants} 
left over from the star formation process (Backman \& Paresce 1993; 
Lagrange, Backman, \& Artymowicz 2000; Zuckerman 2001).

In recent years, $\eri$ has received much attention since it is
thought to be analogous to the solar system
at an age about 10--20\% of that of the Sun:
(1) in comparison with the other three ``Big-Four'' prototypes 
which all are much more luminous A-type stars, 
$\eri$ is {\it closer} in type, mass, and luminosity
($L_\star \approx 0.33 L_\odot$; Soderblom \& D\"appen 1989) 
to the Sun (see Greaves \& Holland 2000);
(2) $\eri$ has a ring-like structure of $\simali$55$\AU$
in radius (which is very similar to the Kuiper Belt in size) 
and a central cavity inside $\simali$30$\AU$ 
(which is equivalent in scale to Neptune's orbit) 
as revealed by its 850$\mum$ image obtained using 
the SCUBA ({\it Submillimetre Common-User Bolometer Array}) 
camera at the {\it James Clerk Maxwell Telescope} 
(JCMT; Greaves et al.\ 1998).
(3) as the oldest (with an age of $\approx 800\Myr$; 
Henry et al.\ 1996) ``Big-Four'' Vega-type star,
the formation of planets in its disk may have completed 
since planet formation is thought to be complete by 
about 10 to 100$\Myr$ (Lissauer 1993);
(4) a roughly Jupiter-mass planet with a period of 
$\simali$6.9$\yr$ appears to be present inside the central 
cleared region of the $\eri$ disk, as suggested by 
the radial velocity measurements spanning 
the years 1980.8--2000.0 (Hatzes et al.\ 2000).
 
To better understand the origin and evolution of protoplanetary 
dust disks and the formation process of planetary systems, 
we need to know the physical, chemical, and dynamical properties 
of their constituent grains.
In previous papers (Li \& Greenberg 1998, Li \& Lunine 2003a,b), 
we have shown that a simple porous dust model is successful 
in reproducing the spectral energy distributions 
from the near-IR to millimeter wavelengths of the disks 
around the Herbig Ae/Be stars $\hra$ and $\hda$
and the Vega-type star $\beta$ Pictoris,
including the 9.7$\mum$ amorphous and the 11.3$\mum$ 
crystalline silicate features or the 7.7$\mum$ and 11.3$\mum$
polycyclic aromatic hydrocarbon (PAH) features.
In this {\it Letter} we will study 
the dust IR to submillimeter emission 
of the disk around the more evolved star $\eri$. 
The objectives of this {\it Letter} are four-fold: 
(1) we wish to infer the physical and chemical
properties of the dust in the $\eri$ disk 
and to infer the disk structure;
(2) we wish to gain insights into the evolution of dust
mass, size, morphology, and mineralogy at different stellar ages; 
(3) we wish to know how widely applicable is the porous dust model;
if it can be shown that the porous dust model is 
``universally'' valid for disks at different evolutionary stages
and with different geometrical structures, it will be 
a valuable guide for interpreting future data sets obtained by 
the upcoming {\it Space Infrared Telescope Facility} (SIRTF); 
(4) we wish to make mid-IR spectral, imaging and broadband 
photometry predictions for the $\eri$ disk; these predictions
can be compared to future SIRTF observations in order to 
test further the validity of the porous dust model. 

\section{The Porous Dust Model\label{sec:model}}
We have discussed in detail in Li \& Greenberg (1998)
and Li \& Lunine (2003a,b) that the dust in protoplanetary
disks most likely consists of porous aggregates of
protostellar materials of interstellar origin, although
there is no consensus regarding the degree to which 
the original interstellar materials have been processed
in protostellar nebulae. Unless there are IR signatures
(e.g. the 11.3$\mum$ crystalline silicate feature)
suggestive of the thermal history of the dust available, 
it is difficult to determine the composition of the dust 
as a priori. Therefore, in modeling the IR emission of dust 
disks we usually invoke two extreme dust types:
``{\it cold-coagulation}'' dust made of coagulated but 
otherwise unaltered protostellar interstellar grains, 
and ``{\it hot-nebula}'' dust made of grains that are 
highly-processed in protostellar nebulae where silicate 
dust is annealed and carbon dust is destroyed by oxidization 
(see Li \& Lunine 2003a,b for details).
Since the $\eri$ disk is too cold to emit at the mineralogical
``fingerprinting'' mid-IR features (e.g. the 9.7 and 11.3$\mum$ 
silicate bands and the 3.3, 6.2, 7.7, 8.6, and 11.3$\mum$ 
PAH bands), we will just adopt the ``cold-coagulation'' dust model.
It has been shown in Li \& Lunine (2003a,b) that
the major differences between the IR emission from these 
two dust models lie in their mid-IR spectral features,
while their continuum emission are very similar.

For the dust properties, we take the porosity 
(or fluffiness; the fractional volume of vacuum in a grain) 
to be $P=0.90$ for 3 reasons:
    (i) dust models with $P\simeq 0.90$ are successful
     in modeling the IR emission from the disks around
     $\hra$ (Li \& Lunine 2003a), $\hda$ (Li \& Lunine 2003b), 
     and $\beta$ Pictoris (Li \& Greenberg 1998);
    (ii) a porosity in the range of $0.80\simlt P\simlt 0.90$
     is expected for dust aggregates formed through coagulation
     as demonstrated both theoretically (Cameron \& Schneck 1965) 
     and experimentally (Blum, Schr\"apler, \& Kozasa 2003);
    (iii) a porosity of $P\simeq 0.90$ for the ``cold-coagulation'' 
     dust is consistent with the mean mass density of cometary nuclei 
     for which the ice-coated ``cold-coagulation'' dust aggregates 
     are plausible building blocks (see Greenberg \& Li 1999).
We assume a power-law dust size distribution 
$dn(a)/da \propto a^{-\alpha}$ which is characterized by 
a lower-cutoff $\amin$, upper-cutoff $\amax$ 
and power-law index $\alpha$;
we take $\amin=1\mum$ and $\amax=1\cm$ 
(see Li \& Lunine 2003a,b).\footnote{%
  We assume all grains are spherical in shape;
  the grain size $a$ is defined as the radius of 
  the sphere encompassing the entire aggregate.
  The choice of $\amin$ and $\amax$, as extensively 
  discussed in Li \& Lunine (2003a,b) for the disks
  around $\hra$ and $\hda$, will not affect our conclusion.   
  }
As far as the dust composition is concerned,
we take the ``cold-coagulation'' dust model --
the dust is assumed to be composed of amorphous silicate 
and carbonaceous materials (and H$_2$O-dominated ices in 
regions colder than $\simali$110--120$\K$);
the mixing mass ratios for the silicate, carbon and 
ice constituent grains are approximately derived in
Li \& Lunine (2003a), by assuming cosmic abundance, 
to be $\mcarb/\msil \approx 0.7$ 
and $\mice/(\msil+\mcarb) \approx 0.8$
where $\msil$, $\mcarb$, and $\mice$ are respectively 
the total mass of the silicate, carbon, and ice subgrains.  

Guided by the 850$\mum$ SCUBA/JCMT image and the 850$\mum$ 
flux radial profile (see Figs.\,1,2 of Greaves et al.\ 1998), 
we will adopt a Gaussian functional formula for the dust 
spatial surface density distribution
$\sigmar = \sigmap \exp[-4\ln2\{(r-\rp)/\Delta\}^2]$
which describes the dust ring peaking at $\simali$55$\AU$ 
from the star.
The extent of the $\eri$ disk is cut off at 
an inner boundary of $\rin =0.05\AU$ where small grains
are heated to temperatures $\gtsim 1500\K$,
and an outer boundary of $\rout=200\AU$ outside of which
there is little dust emission.

We use Mie theory together with the Bruggman effective 
medium theory (Kr\"ugel 2003) to calculate the absorption 
cross sections of the fluffy heterogeneous dust aggregates. 
Dielectric functions are taken from 
Draine \& Lee (1984) for amorphous silicate dust;
Li \& Greenberg (1997) for carbonaceous dust; and
Li \& Greenberg (1998) for H$_2$O-dominated ``dirty'' ice.
The $\eri$ stellar radiation is approximated by
the Kurucz (1979) model atmosphere spectrum for K2V stars 
with an effective temperature of $\Teff=5000\K$, 
a surface gravity of ${\rm lg}$\,$g$=4.0, 
and a solar metallicity $Z/Z_\odot=1$.
The calculated dust IR emission spectrum is compared 
with the IRAS flux at 12, 25, 60, and 100$\mum$ 
and the SCUBA flux at 450 and 850$\mum$ 
(see Table 1 in Greaves et al.\ 1998).
There also exist 1300$\mum$ photometric measurements
for the $\eri$ disk 
(e.g., Chini, Kr\"ugel, \& Kreysa 1990, Chini et al.\ 1991),
but we will not include these data since these measurements 
were all made using single-element bolometers with telescope 
beams smaller than the disk extent so that the flux obtained
by these measurements may be underestimated 
(Zuckerman \& Becklin 1993, Weintraub \& Stein 1994).

\section{Model Results\label{sec:results}}
The model has four free parameters: 
$\alpha$ -- the dust size distribution power-law exponent,
$\rp$ -- the disk radial location where the dust spatial distribution 
peaks; $\Delta$ -- the FWHM of the dust spatial distribution;
$\sigmap$ -- the mid-plane surface density at $r=\rp$.
Observational constraints include the observed IR spectral energy
distribution and the SCUBA 850$\mum$ image. The latter is particularly
important for constraining the spatial distribution of the dust in
the $\eri$ disk. Instead of using the smoothed 850$\mum$ image 
presented in Greaves et al.\ (1998), we constructed an unsmoothed 
850$\mum$ image from all 1997--2000 data in the SCUBA archive 
using the data reduction procedure outlined in Bendo et al.\ (2003).
However, both the image and the radial profile obtained 
from our analysis show an overall good agreement with 
those of Greaves et al.\ (1998).

\begin{figure}[h]
\begin{center}
\epsfig{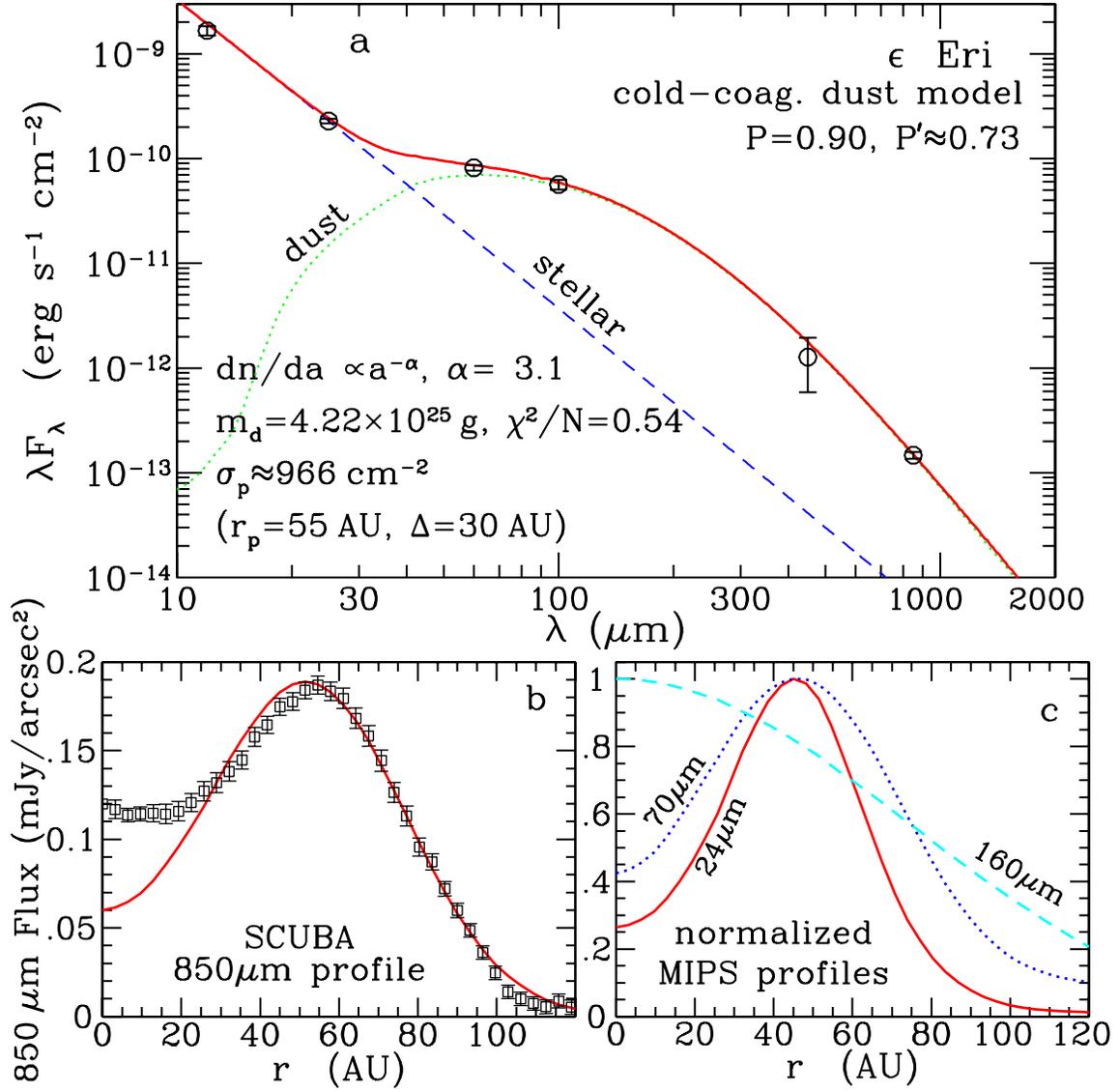}
\end{center}\vspace*{-1em}
\caption{
        \label{fig:sed}
        \footnotesize
        {\bf (a)} IR to submillimeter emission of the $\eri$ system
        (open circles: the 12, 25, 60, and 100$\mum$ IRAS data
         and the 450 and 850$\mum$ SCUBA data) compared with 
         the model spectrum (solid line; which is the sum of 
         the stellar [dashed line] and the dust [dotted line] 
         contributions).
        {\bf (b)} Radial profile of the dust emission at 850$\mum$.
         Open squares plot the azimuthally averaged flux density 
         radial distribution calculated from the SCUBA 850$\mum$ image 
         (also see Figs.\,1,2 of Greaves et al.\ 1998 but theirs
          were smoothed with an $8^{\prime\prime}$ Gaussian).
          The apparent rise within $r\approx 15\AU$
          is very likely artificial (see \S\ref{sec:results}).
          Solid line plots the model-predicted 
          850$\mum$ emission radial profile
          convolved with the SCUBA 14$^{\prime\prime}$ beam. 
        {\bf (c)} Radial profiles predicted for the SIRTF MIPS bands
         at 24$\mum$ (solid line), 70$\mum$ (dotted line), 
         and 160$\mum$ (dashed line). All profiles are convolved 
         with the MIPS model PSFs and normalized to their peak values: 
         $\approx 0.71$, 6.47, and 6.12\,${\rm mJy\,arcsec^{-2}}$ 
         for the 24, 70, and 160$\mum$ bands, respectively.
        }
\end{figure}

As shown in Figure \ref{fig:sed}, 
the $P=0.90$ porous dust model\footnote{%
 In fact the porosity is reduced to $\Pice \approx 0.73$ 
 after ices fill in some of the vacuum 
 (see Appendix B in Li \& Lunine 2003a)
 which occurs at $r\simgt 3\AU$.
 }
with $\alpha \approx 3.1$, $\rp\approx 55\AU$, $\Delta \approx 30\AU$, 
and $\sigmap \approx 966\cm^{-2}$ provides an almost perfect 
fit to the observed IR to submillimeter emission (Fig.\,\ref{fig:sed}a)
and to the azimuthally averaged flux radial distribution at 850$\mum$
(Fig.\,\ref{fig:sed}b) except that the model is deficient  
in emitting at 850$\mum$ inside $\simali$28$\AU$ (see below for
further discussion).\footnote{%
  The model-predicted 850$\mum$ radial profile is convolved
  with the SCUBA $14^{\prime\prime}$ beam using 
  the actual SCUBA point spread function (PSF).   
  }
This model requires a total dust mass of 
$\md\approx 4.22\times 10^{25}\g \approx 7.1\times 10^{-3}\mearth$.
Let the goodness of fit be $\chisq \equiv 
\frac{1}{N_{\rm obs}} \sum_{i=1}^{N_{\rm obs}}
\left\{ \left( \left[\lambda F_{\lambda}\right]_{\rm mod}
- \left[\lambda F_{\lambda}\right]_{\rm obs}\right)
/\left[\lambda \Delta F_{\lambda}\right]_{\rm obs} \right\}^2$
where $\left[\lambda F_{\lambda}\right]_{\rm obs}$ and 
$\left[\lambda F_{\lambda}\right]_{\rm mod}$ are respectively
the observed and model-predicted flux densities, 
$\left[\lambda \Delta F_{\lambda}\right]_{\rm obs}$ 
is the observational flux uncertainty, $N_{\rm obs}\equiv 6$ is 
the number of data points to be fit. 
This model gives $\chisq\approx 0.54$.

The apparent rise in flux density within 
$r$$\approx$$4.5^{\prime\prime}$ ($\simali$15$\AU$) 
was also seen in the smoothed SCUBA 850$\mum$ 
radial profile of Greaves et al.\ (1998; see their Fig.\,2).
This rise may be associated with the 
``{\it small peak $5^{\prime\prime}$--$10^{\prime\prime}$ south of star}'' 
seen in the SCUBA 850$\mum$ image (see Fig.\,1 of Greaves et al.\ 1998).
If this is real, we then need to resort to an additional dust component
lying close to the star. This inner warm component would
resemble the zodiacal dust cloud in our solar system.
However, the SCUBA images are quite noisy in the inner region, so the
anomalous emission inside $r\approx 4.5^{\prime\prime}$ may be 
instrumental noise or image processing artifacts.  
We therefore exclude this ``inner component'' emission 
in our model.

\section{Discussion\label{sec:discussion}}
We have seen in \S\ref{sec:results} that the very same porous 
dust model shown to be successful in modeling the IR emission
from the dust disks around the Herbig Ae/Be stars
$\hra$ (Li \& Lunine 2003a) and $\hda$ (Li \& Lunine 2003b)    
and the young MS star $\beta$ Pictoris (Li \& Greenberg 1998)
is also successful in reproducing the IR to submillimeter SED 
as well as the 850$\mum$ flux density radial profile 
of the more evolved MS star $\eri$ ($\simali$$800\Myr$ old),
although the model appears unable to provide
sufficient 850$\mum$ emission inside $\simali$28$\AU$ 
(see Fig.\,\ref{fig:sed}b). It is very likely that 
this inner component emission is not real (see \S\ref{sec:results});
but if one really wants to push us to account for this emission,
we find that it could be explained by invoking
an inner ``zodiacal'' dust cloud of 
$\simali$$5.2\times 10^{-4}\mearth$ mass
located at $\simali$2$\AU$ from the star. 

SIRTF will be capable of sensitive imaging using 
the {\it Infrared Array Camera} (IRAC) 
at 3.6, 4.5, 5.8, and 8.0$\mum$, 
and using the {\it Multiband Imaging Photometer} (MIPS) 
at 24, 70, and 160$\mum$.
SIRTF will also be able to perform low-resolution 5--40$\mum$
and high-resolution 10--37$\mum$ spectroscopic observations 
using the {\it Infrared Spectrograph} (IRS) instrument.
However, the dust in the $\eri$ disk seems to be too cold
to emit in an appreciable quantity in the IRAC bands
(see Fig.\,\ref{fig:sed}a).
Neither can this dust be heated sufficiently to emit
at the characteristic mid-IR spectral features.
But the MIPS imaging will provide powerful constraints 
on the $\eri$ dust spatial distribution. We have therefore
calculated the MIPS 24, 70, and 160$\mum$ 
flux density radial distributions for the best-fit porous 
dust model (see Fig.\,\ref{fig:sed}c).
All profiles are convolved with the MIPS model PSFs.
In Table \ref{tab:sirtf} we show the model-predicted
band-averaged intensities.

\begin{table}[h]
\begin{center}
\caption[]{Dust IR emission (Jy) integrated over 
           SIRTF bands predicted for the porous dust model. 
           Also tabulated are the stellar photospheric 
           emission (Jy)\label{tab:sirtf}.}
\begin{tabular}{cccccccc}
\hline
Instrument & IRAC & IRAC & IRAC & IRAC & MIPS  & MIPS &MIPS \\
$\lambda_{\rm eff}$ & 3.6$\mum$ & 4.5$\mum$ & 5.8$\mum$ & 8$\mum$ 
          & 24$\mum$ & 70$\mum$ & 160$\mum$ \\
\hline
dust    & $1.35\times 10^{-5}$ & $2.03\times 10^{-5}$ 
        & $3.43\times 10^{-5}$ & $8.47\times 10^{-5}$
        & 0.10 &1.55 &1.71\\
stellar & 65.4 & 40.7 
        & 28.6 & 16.4
        & 2.26 & 0.25 & 0.056\\
\hline
\end{tabular}
\end{center}
\end{table}

\begin{figure}[]
\begin{center}
\epsfig{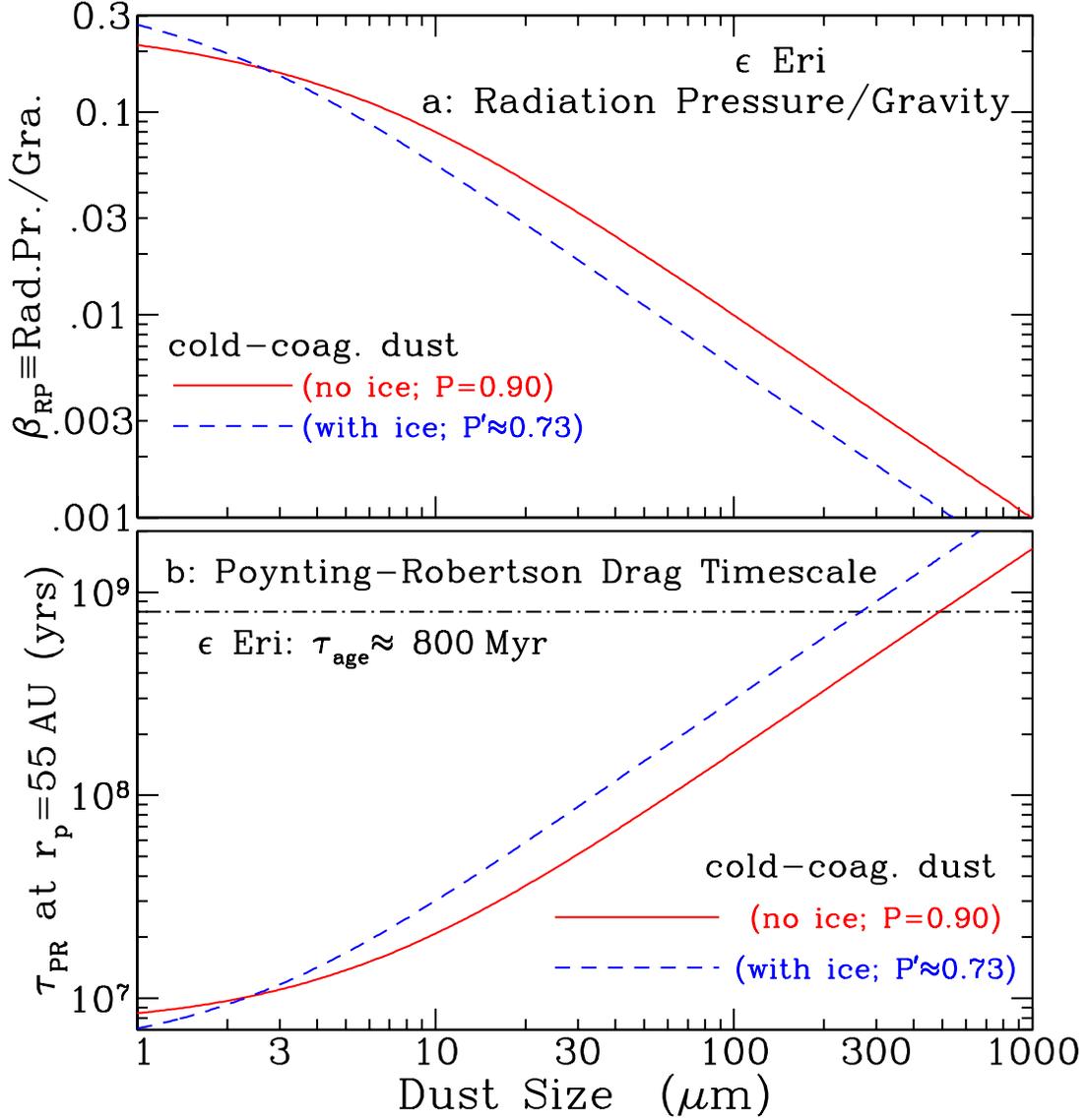}
\end{center}\vspace*{-1em}
\caption{
        \label{fig:rppr}
        \footnotesize
        {\bf (a)} Ratio of the radiative repulsion to 
        the gravitational attraction ($\beta_{\rm RP}$) for 
        the best-fit ``cold-coagulation'' dust 
        (without ice $P=0.90$ [solid line] 
        or with ice $\Pice=0.73$ [dashed line]).
        {\bf (b)} The orbit decay timescales $\tau_{\rm PR}$
        due to the Poynting-Robertson drag for 
        the best-fit ``cold-coagulation'' dust
        (without ice $P=0.90$ [solid line] 
        or with ice $\Pice=0.73$ [dashed line])
        at a radial distance of $\rp=55\AU$ from the central star
        (note $\tau_{\rm PR}\propto r^{2}$: 
         the PR timescale decreases for dust 
         at a smaller radial distance).
        The dot-dashed horizontal line plots 
        the $\eri$ age ($\approx 800 \Myr$).  
        }
\end{figure}

The dust in the $\eri$ disk is stable against the radiation
pressure. As shown in Figure \ref{fig:rppr}a,
$\beta_{\rm RP}$, the ratio of radiative pressure (RP) force 
to gravitational force is found to be smaller than 0.3 
for all grains. But the Poynting-Robertson (PR) drag would
remove grains smaller than $\simali$270$\mum$ at $r = 55\AU$ 
in a timescale ($\tau_{\rm PR}$) shorter than the age of 
the $\eri$ system (see Fig.\,\ref{fig:rppr}b).\footnote{%
 {\it Closer} to the star, the threshold grain size 
 below which grains will be ejected {\it increases} 
 since $\tau_{\rm PR} \propto r^2$ 
 (see Eq.22 in Li \& Lunine 2003b).
 }
By integrating the PR dust removal rate
$\left(4\pi/3\right)\rho a^3/\tau_{\rm PR}(a,r)$ 
[where $\rho$ is the mass density of the dust aggregate:
$\rho \approx 0.25\,(0.46) \g\cm^{-3}$ for the $P=0.90$ 
($\Pice=0.73$ ice-coated) cold-coagulation type dust;
see Appendix B in Li \& Lunine 2003a] 
over the whole size range and over the entire disk, 
we estimate the PR dust mass loss rate to be 
$\approx 4.3\times 10^{-11}\,m_\oplus\yr^{-1}$ 
for the porous dust model.
Over the life span of $\eri$, roughly $0.034\,m_\oplus$ 
of dust is lost by the PR drag.
Therefore, the observed grains need to be continuously replenished,
very likely from cascade collisions and evaporation of larger bodies 
such as comets and/or asteroids.

The inner ``zodiacal'' cloud, {\it if it indeed exists},
could be maintained by grains released from the sublimation of 
bombarding comets near the star.\footnote{%
  In contrast, the cold outer ring is more likely
  to be collisionally replenished, although the sublimation 
  of cometary CO volatiles (with a sublimation temperature of
  $\simali$20$\K$) occurring inside $r$$<$110$\AU$ does generate 
  dust outflow. But dust outflow only becomes significant 
  when comets are close to the star, say, at $r\simlt 3\AU$ 
  where H$_2$O ice starts to sublimate.   
 }
Taking the total dust mass loss $\simali$$3\times 10^{16}\g$ 
in the 1997 apparition of comet Hale-Bopp (Jewitt \& Matthews 1999)
as a representative dust supply rate (per comet per apparition), 
we estimate a cometary infalling rate of $\simali$7$\yr^{-1}$, 
roughly consistent with the estimation of $<$15$\yr^{-1}$ 
derived by Dent et al.\ (1995) from the observed CO upper limit 
assuming sublimating comets as a CO replenishment source 
in the $\eri$ disk.\footnote{%
  If we adopt the Halley dust loss rate $\simali$$1\times 10^{14}\g$
  per orbit (Lisse 2002), we would expect a comet infall rate
  $\simali$300 times more frequent. This is much higher than
  what is observed ($\simali$20--30 comets per year)
  for the present solar system.
  But we note that the number of cometary bodies and 
  the bombardment rate were much higher at the early stage
  of our solar system.
  The solar system zodiacal cloud has a mass loss rate
  of $\simali$$5\times10^{-14}\mearth \yr^{-1}$;
  Lisse (2002) argued that the short-period comets
  alone are enough to supply the dust.
  In the $\beta$ Pictoris disk, the number of {\it observed}
  cometary bodies (crossing the line of sight) is 
  $\simali$300--400 per year (Beust \& Morbidelli 1996);
  however, the actual number would be much higher since
  many objects do not cross the line of sight
  and are therefore not visible in spectroscopy.
  }

The grain-grain collision timescale is only $\approx 0.1\Myr$
for dust at $\rp=55\AU$, significantly shorter
than the PR timescale $\tau_{\rm PR}$.
But it is unlikely for grain-grain collisions to 
effectively remove the dust from the disk; instead, 
their major role is to re-distribute the dust over 
different size bins through fragmentation.
The partially cleared central cavity is most likely created 
by the accumulation of gas and grains into planetesimals,
as supported by the non-detection of 
circumstellar gas in this disk (Dent et al.\ 1995, Liseau 1999);
and by the fact that the PR drag alone is not sufficient 
to clear the dust out to large radii up to $\simali$30$\AU$
(Jura 1990, Greaves et al.\ 1998, Dent et al.\ 2000).

Dent et al.\ (2000) modeled the $\eri$ SED 
in terms of {\it compact} grains, 
assuming the dust absorption efficiency to be
$Q_{\rm abs} =1$ for $\lambda$$<$$30\mum$
and $Q_{\rm abs} =\left(\lambda/30\mum\right)^{-0.8}$ 
for $\lambda$$\ge$$30$$\mum$,
and adopting a uniform surface density distribution 
[i.e. $\sigmar \equiv {\rm constant}$] in the range
of 50$\simlt$$r$$\simlt$80$\AU$. Satisfactory fit to 
the observed SED was achievable.
Using emissivities calculated from real dust mixtures,
however, Sheret, Dent, \& Wyatt (2003) found that compact 
grains are not able to fit the observed SED.

The model presented in this {\it Letter} is somewhat simplified 
in the sense that we have assumed a symmetrical morphology 
for the $\eri$ disk. Asymmetries and discrete flux enhancements 
(``clumps'' or ``bright spots'') are actually
seen in its 850$\mum$ SCUBA images (Greaves et al.\ 1998). 
As suggested by Ozernoy et al.\ (2000) 
and Quillen \& Thorndike (2002), these features may be caused
by the capture of grains into the mean motion resonances
with a planet of Jupiter-mass embedded in the disk.
This simplification would not affect the general conclusions 
of this {\it Letter} since the IR emission is mainly determined by 
the general distribution of dust in the disk, that is, 
a narrow ring. 

\section{Conclusion\label{sec:conclusion}}
The porous dust model, previously developed 
for the extended wedge-like disk around 
the $\simali$15$\Myr$-old MS star $\beta$ Pictoris 
and the narrowly confined ring-like disks around 
the Herbig Ae/Be stars $\hra$ ($\simali$8$\Myr$ old) 
and $\hda$ ($\simali$5$\Myr$ old), is shown also applicable
to the dust disk around the $\simali$800$\Myr$-old MS star $\eri$,
the oldest ``Big-Four'' Vega-excess star.
Modeled as a ring peaking at $\simali$55$\AU$ from
the star with a FWHM $\simali$30$\AU$,
analogous to the young solar system Kuiper Belt,
the observed IR to submillimeter emission 
as well as the 850$\mum$ flux density radial 
distribution of the $\eri$ disk is closely reproduced
by porous grains of $\simali$$7.1\times 10^{-3}\mearth$ mass
consisting of $\simali$90\% vacuum in volume.
Predictions are made for future SIRTF/MIPS imaging observations
that will offer a useful test of the porous dust model. 

\acknowledgments
We thank 
J.S. Greaves,
E.K. Holmes,
A. Lecavelier des Etangs, 
R. Malhotra, 
K.A. Misselt,
C. Papovich,
W.T. Reach, 
G. Schneider, 
Z. Sekanina,
M.D. Silverstone,
and the anonymous referee for 
helpful discussions and/or suggestions.
A. Li thanks the University of Arizona for the ``Arizona 
Prize Postdoctoral Fellowship in Theoretical Astrophysics''.
This research was supported in part by a grant from the NASA 
origins research and analysis program.
G.J. Bendo is a Guest User of Canadian Astronomy Data Centre, 
which is operated by the Dominion Astrophysical Observatory 
for the National Research Council of Canada's Herzberg Institute 
of Astrophysics.

\end{document}